\documentclass[pdflatex,sn-nature]{sn-jnl}

\usepackage[T1]{fontenc}
\usepackage{graphicx,amsmath,amssymb,amsfonts,mathrsfs}
\usepackage{booktabs,array,longtable,calc,multirow}
\usepackage{adjustbox,placeins}
\usepackage{textcomp}

\begin{document}
\title{
Microscopic constitutive freedom in metamaterials}
\author[1,†]{Zhongjian Mei}
\author[2,†]{Chuanhao Ge}
\author[2]{Nengyin Wang}
\author[3]{J. Stuart Bolton}
\author[1,4,*]{Tongyang Shi}
\author[2,*]{Yong Li}
\author[1,4,5,*]{Jun Yang}

\affil[1]{Institute of Acoustics, Chinese Academy of Sciences, Beijing 100190, China}
\affil[2]{Institute of Acoustics, Tongji University, Shanghai 200092, China}
\affil[3]{Ray W. Herrick Laboratories, Purdue University, West Lafayette, Indiana 47907, USA}
\affil[4]{State Key Laboratory of Acoustics and Marine Information, Institute of Acoustics, Chinese Academy of Sciences, Beijing 100190, China}
\affil[5]{University of Chinese Academy of Sciences, Beijing 100049, China}
\begingroup
\renewcommand{\thefootnote}{}
\footnotetext{$^\dag$ These authors contributed equally to this work.}
\footnotetext{$^*$ e-mail: shitongyang@mail.ioa.ac.cn; yongli@tongji.edu.cn; jyang@mail.ioa.ac.cn}
\endgroup

\abstract{
Metamaterials expand the range of accessible macroscopic responses through architectural design. Here, a cross-scale design strategy is introduced that combines microscopic control of static constitutive properties with spectral engineering through macroscopic architecture. Beyond shaping frequency-dependent responses, this strategy modifies the corresponding broadband performance bounds in passive systems by engineering the static parameters entering causal constraints. The concept is realized in granular activated carbon, where reversible sorption enhances equilibrium compressibility. Extrapolation of a multiscale model supported by low-frequency measurements to the zero-frequency limit yields an effective static bulk modulus approximately half the conventional nonsorbing isothermal reference value. This reduction substantially lowers the corresponding causal minimum-thickness bound for rigidly backed absorbers. A cascaded hierarchical composite architecture further redistributes absorption towards the target frequency band, achieving near-perfect absorption with a normal-incidence absorption coefficient above 0.9 across most of the 40–6400 Hz range. Together, static constitutive control and macroscopic spectral design enable ultrabroadband sound absorption beyond the conventional isothermal thickness benchmark without violating causality. This strategy offers a new approach to extending the attainable performance of metamaterials.
}
\maketitle

\section{Introduction}

Macroscopic material properties emerge from microscopic structure and interactions, whose collective response to external perturbations is captured by constitutive relations.
Metamaterials have expanded the accessible material-property landscape by introducing artificial architecture as an additional degree of freedom, enabling effective responses that are difficult to obtain from conventional media\cite{Kadic-2019-NRP}. Geometry, connectivity and internal organization can therefore be engineered to tailor macroscopic behaviour across electromagnetics and optics\cite{Shelby-2001-S, Yu-2011-S}, mechanics\cite{Djellouli-2024-N, Zheng-2014-S}, thermal science\cite{Schittny-2013-PRL} and acoustics\cite{Cummer-2016-NRM,Ge-2018-NSR}.


In acoustics, subwavelength architectures provide extensive freedom to engineer effective density, bulk modulus, impedance and frequency-dependent response\cite{Fang-2006-NM,Ding-2007-PRL,Yang-2008-PRL,Assouar-2018-NRM}. Local resonances\cite{Liu-2000-S}, coiled channels\cite{Liang-2012-PRL,Li-2016-APL} and coupled modes\cite{Ma-2014-NM} enable versatile control over wave propagation, scattering and absorption, while resonance, dispersion and modal engineering allow acoustic responses to be redistributed and shaped across frequency\cite{Huang-2020-SB,Zhu-2021-PRB,Wang-2025-NSR}. Such developments have made spectral response an increasingly accessible dimension of acoustic material design\cite{Ma-2016-SA}.


Yet, the frequency-dependent properties of a material are not independent design variables. In passive, linear and time-invariant systems, causality and passivity connect spectral behaviour to the material’s static or asymptotic constitutive properties\cite{Gustafsson-2010-NJP,Norris-2018-JASA,Meng-2022-PRSA}. Metamaterial architectures can redistribute and reshape responses across frequency, but the attainable broadband behaviour remains tied to this underlying constitutive reference\cite{Yang-2017-MH,Qu-2025-NC,Qu-2026-PRL}. This relation suggests a complementary direction for material design: engineering not only the frequency dependence of a response, but also the constitutive properties that fundamentally constrain it\cite{Yang-2025-CP,Zhang-2026-NPJA}.


Here, we pursue this constitutive route using reversible sorption in granular activated carbon (GAC) to engineer the static acoustic response\cite{Bechwati-2012-JASA,Venegas-2016-JASA}. A multiscale theoretical framework connects sorption to equilibrium compressibility and the effective static bulk modulus\cite{Venegas-2017-PF}, while extrapolation of low-frequency measurements to the zero-frequency limit yields a normalized static bulk modulus of 0.567 relative to the conventional isothermal reference. Incorporating this sorption-modified modulus into the causal thickness relation predicts a substantially reduced minimum thickness for rigidly backed absorbers. We further combine this static constitutive control with structural and spectral engineering in a graded composite comprising GAC, melamine foam and air channels. The resulting absorber achieves efficient broadband absorption over 40–6400 Hz with a physical thickness of 350 mm, compared with a conventional isothermal causal thickness benchmark of 538 mm for the same target spectrum, corresponding to a reduction of approximately 35\%. Together, these results expand acoustic metamaterial design beyond frequency-dependent response engineering to include the static constitutive parameters that underlie broadband performance constraints.

\FloatBarrier
\section{Molecular control of acoustic constitutive response}
\begin{figure*}[!htbp]\centering
\includegraphics[width=138mm]{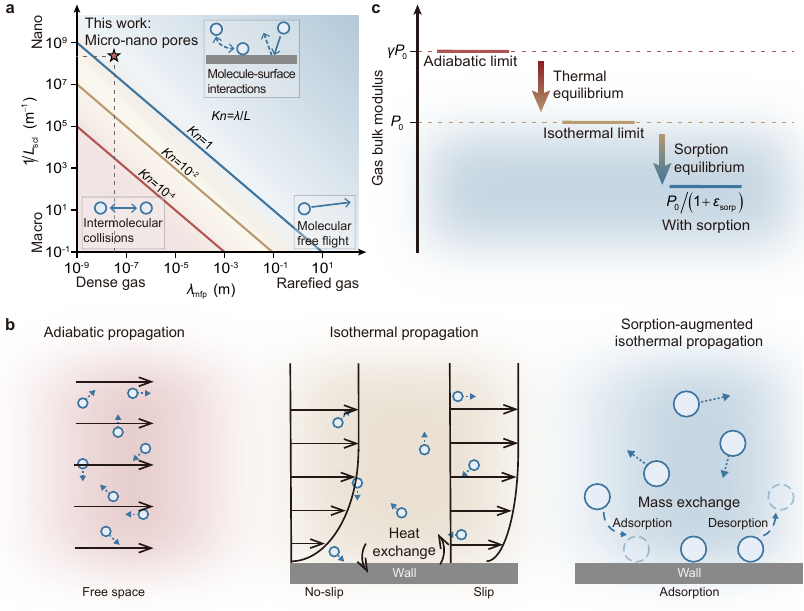}
\caption{\textbf{Microscopic processes and acoustic response of a gas.} (a) Schematic map of intermolecular collisions, molecular free flight and molecule--surface interactions as functions of the molecular mean free path, $\lambda_{\mathrm{mfp}}$, and inverse characteristic structural size, $1/L_{\mathrm{scl}}$. Diagonal lines mark constant Knudsen number, $\mathrm{Kn}=\lambda_{\mathrm{mfp}}/L_{\mathrm{scl}}$; the star indicates the micro- and nanoscale pore region examined here. (b) Approximately adiabatic propagation in free space, wall-mediated heat exchange and no-slip or slip velocity profiles under confinement, and reversible adsorption and desorption at a sorptive surface. (c) Gas bulk modulus under adiabatic compression ($\gamma P_0$), isothermal compression ($P_0$) and sorption equilibrium [$P_0/(1+\varepsilon_{\mathrm{sorp}})$]. Here, $P_0$ is the equilibrium pressure, $\gamma$ is the heat-capacity ratio and $\varepsilon_{\mathrm{sorp}}$ is the sorption-induced enhancement of equilibrium compressibility relative to the nonsorbing isothermal gas.}
\label{fig:1}
\end{figure*}

The acoustic response of a gas emerges from the collective dynamics of its molecules. Under ordinary atmospheric conditions and at conventional acoustic scales, intermolecular collisions are sufficiently frequent to sustain approximate local thermodynamic equilibrium, providing the basis for a continuum description of sound propagation\cite{Shen-2005-Book}. As the molecular mean free path increases, molecules travel farther between collisions, and transport over a given distance acquires a greater free-flight or ballistic character. A complementary change occurs under geometric confinement: even at an unchanged mean free path, reducing the characteristic structural dimension increases the relative frequency of molecular encounters with solid boundaries. Molecule–surface interactions consequently become increasingly important in micro- and nanoscale structures. Such interactions involve momentum and energy exchange and, for sorptive surfaces, may additionally include adsorption and desorption\cite{Umnova-2009-JASA,Sam-2023-JPCC,Didier-2025-PRM}. Thus, the microscopic processes underlying a macroscopic acoustic response are not fixed, but depend on the relative molecular and structural length scales, ranging from collision-dominated dynamics to regimes in which molecular free flight or gas–surface interactions become increasingly important. The relative importance of these processes can be related to the molecular and structural length scales through the Knudsen number, $\mathrm{Kn}=\lambda_{\mathrm{mfp}}/L_{\mathrm{scl}}$, where \(\lambda_{\mathrm{mfp}}\) is the molecular mean free path and \(L_{\mathrm{scl}}\) is the characteristic structural size (Fig.~\ref{fig:1}(a)). In this work, we focus on the confinement route, examining how molecule–surface processes that become significant in micro- and nanoscale structures modify the macroscopic acoustic response.

In free space, acoustic compression and expansion of a gas are approximately adiabatic and accompanied by temperature fluctuations. Geometric confinement introduces wall-mediated heat exchange, which relaxes these fluctuations and drives the compression response toward the isothermal limit as thermal equilibration becomes sufficiently rapid relative to the acoustic period\cite{Allard-2009-Book,Champoux-1991-JAP}. The same molecule–wall interactions also transfer momentum, shaping the near-wall velocity field and giving rise to no-slip or velocity-slip behaviour at the boundary\cite{Maxwell-1879-PTRSLA,Umnova-2009-JASA}. As the characteristic structural dimension decreases further into the micro- and nanoscale, molecule–surface interactions become increasingly prominent and, for surfaces with appreciable affinity for gas molecules, can additionally involve adsorption and desorption. Because the equilibrium amount of adsorbed gas depends on pressure, acoustic pressure oscillations drive reversible molecular redistribution between the gas and adsorbed phases: increasing pressure promotes adsorption, whereas decreasing pressure promotes desorption\cite{Nori-2014-JASA,Venegas-2017-WM}. Sorption therefore modifies the conventional isothermal compression response by introducing an additional pressure-dependent molecular redistribution between the two phases (Fig.~\ref{fig:1}(b)).

These thermodynamic responses are directly reflected in the bulk modulus of the gas. Under adiabatic compression, pressure fluctuations are accompanied by temperature variations, giving the familiar modulus \(B_{\mathrm{ad}}=\gamma P_0\). As thermal equilibration suppresses these temperature fluctuations, the compression response approaches the isothermal limit, with \(B_{\mathrm{iso}}=P_0\), which provides the conventional static reference for a nonsorbing gas. Reversible sorption introduces an additional contribution to the equilibrium compressibility: a pressure perturbation not only compresses the free gas but also changes the equilibrium amount of adsorbed molecules. This pressure-dependent redistribution provides an additional degree of freedom for accommodating compression, thereby increasing the equilibrium compressibility\cite{Bechwati-2012-JASA,Venegas-2016-JASA,Venegas-2017-PF}. For an idealized gas–surface system, the corresponding sorption-modified static bulk modulus is

\begin{equation}
B_{\mathrm{sorp},0} = \frac{P_0}{1+\varepsilon_{\mathrm{sorp}}}, 
\end{equation}
where \(\varepsilon_{\mathrm{sorp}}\) quantifies the sorption-induced enhancement of the equilibrium compressibility relative to the nonsorbing isothermal reference. Within the adopted Langmuir description\cite{Langmuir-1916-JACS,Do-1998-Book}, \(\varepsilon_{\mathrm{sorp}}=(\rho_N/\rho_0)P_0b/(1+P_0b)^2\), combining the relative sorption capacity with the pressure sensitivity of equilibrium adsorption (details of the idealization and derivations are provided in the Supplementary Material, Sec. I and III). Reversible sorption therefore increases the equilibrium compressibility beyond the conventional isothermal value, reducing the static bulk modulus below \(P_0\) (Fig.~\ref{fig:1}(c)).

The significance of this reduced static modulus becomes evident through its role in the causal constraint on broadband sound absorption\cite{Rozanov-2000-ITAP,Yang-2017-MH}. For a passive, rigidly backed absorber, the causality-based minimum thickness can be expressed as\cite{Yang-2017-MH}
\begin{equation}
L_{\min}
=
\frac{1}{4\pi^{2}}
\frac{B_{\mathrm{eff}}}{B_{0}}
\left|
\int_{0}^{\infty}
\ln\!\left[1-A(\lambda)\right]
\,\mathrm{d}\lambda
\right|,
\end{equation}
where $B_{\mathrm{eff}}$ is the effective static bulk modulus of the material, $B_{0}$ is the bulk modulus of the surrounding medium, and $A(\lambda)$ is the wavelength-dependent absorption spectrum. For a prescribed absorption spectrum in a fixed surrounding medium, the spectral integral and $B_{0}$ are fixed, such that a reduction in $B_{\mathrm{eff}}$ directly reduces the causality-based minimum thickness. Previous causality-based analyses based on the conventional nonsorbing isothermal constitutive reference have therefore established a corresponding isothermal thickness benchmark\cite{Ge-2025-PRL}. Sorption extends this picture by reducing $B_{\mathrm{eff}}$ below that reference, thereby lowering the minimum thickness required for the same absorption target. This constitutive reduction provides a route to thinner broadband absorbers by modifying the static material parameter entering the causal thickness constraint.

\FloatBarrier
\section{Sorption-modified bulk modulus and sound absorption in GAC}
\begin{figure}[h]\centering
\includegraphics[width=138mm,keepaspectratio]{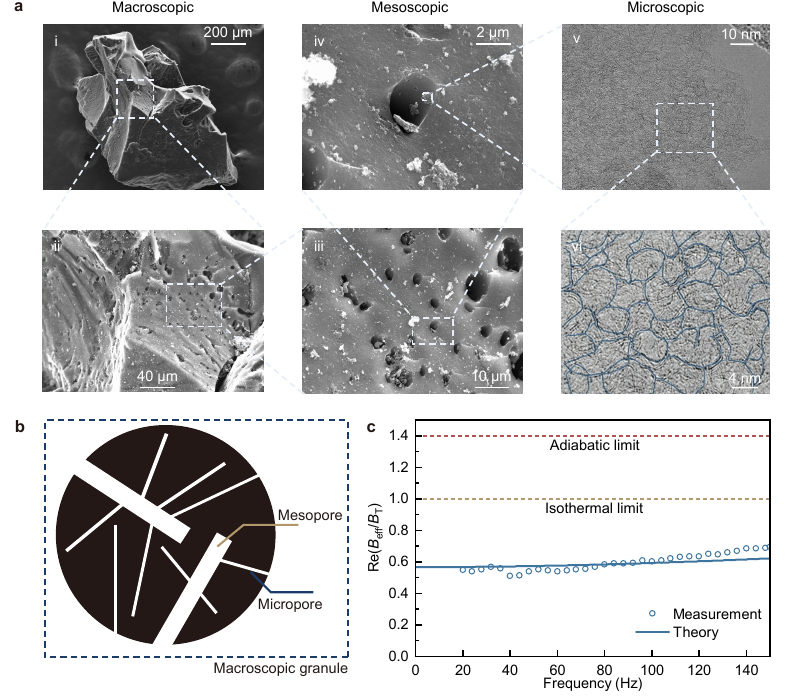}
\caption{\textbf{Multiscale pore structure and sorption-modified effective bulk modulus of granular activated carbon.}
(a) Scanning electron microscopy images (i--iv) and transmission electron microscopy images (v, vi) of granular activated carbon (GAC), showing the particle morphology, intragranular pores, and nanoscale structure. Dashed boxes and connecting lines indicate corresponding regions across scales, while blue lines highlight nanoscale structural contours.
(b) Schematic of an activated carbon granule containing mesopores and micropores.
(c) Frequency dependence of the real part of the normalized effective bulk modulus, $\mathrm{Re}(B_{\mathrm{eff}}/B_{\mathrm{T}})$, where $B_{\mathrm{eff}}$ is the effective bulk modulus of GAC and $B_{\mathrm{T}}=P_0$ is the conventional isothermal reference modulus, with $P_0$ denoting the equilibrium pressure. Open circles show measurements obtained using the two-thickness method, and the blue solid line shows the theoretical prediction including sorption. Red and brown dashed lines mark the adiabatic and conventional isothermal reference levels, respectively.}
\label{fig:2}
\end{figure}

Realizing the sorption-modified constitutive response predicted above requires a porous material with abundant internal gas--solid interfaces and appreciable adsorption capacity\cite{Venegas-2016-JASA,Didier-2025-PRM}. To maximize this interfacial contribution, we selected PCK14E granular activated carbon (GAC), which combines a hierarchical pore structure with a high specific surface area of $1500~\mathrm{m^2\,g^{-1}}$. Scanning electron microscopy (SEM) and transmission electron microscopy (TEM) reveal a multiscale porous morphology encompassing intergranular voids, intragranular pores, and nanoscale features (Fig.~\ref{fig:2}(a)). This hierarchical structure provides extensive internal interfaces for gas sorption and is represented by the triple-porosity model linking the local sorption response to the macroscopic constitutive behaviour\cite{Venegas-2016-JASA,Venegas-2017-PF} (Fig.~\ref{fig:2}(b); Supplemental Material, Sec. II).

To assess whether this structure realizes the predicted constitutive response, we determined the frequency-dependent effective bulk modulus of GAC using the two-thickness method. Across the measured frequency range, the experimental results agree well with the sorption-inclusive theoretical prediction and remain below the conventional isothermal reference (Fig.~\ref{fig:2}(c); see Supplemental Material, Section IV for experimental measurement method of effective static bulk modulus). Extrapolation of the validated model to the zero-frequency limit yields a normalized effective static bulk modulus of
$B_{\mathrm{eff},0}/{B_T}=0.567$,
where $B_T=P_0$ is the conventional isothermal reference. These results support the realization in GAC of a sorption-modified static constitutive response below the conventional isothermal reference.

\begin{figure}[!htbp]\centering
\includegraphics[width=\linewidth,keepaspectratio]{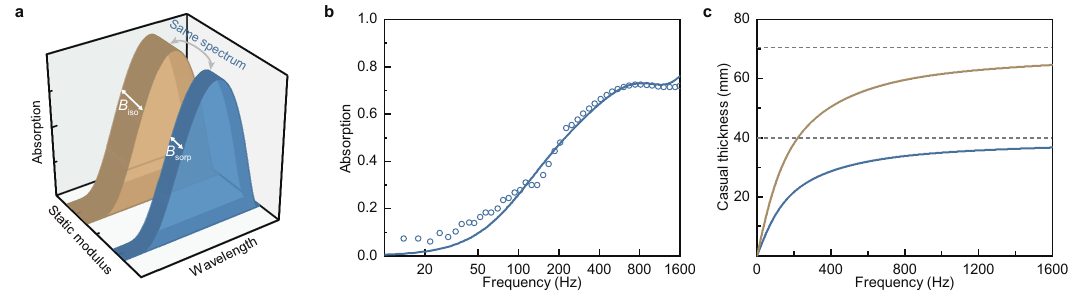}
\caption{\textbf{Static-modulus control of the thickness required for sound absorption.}
(a) Three-dimensional schematic with axes representing wavelength, absorption, and static bulk modulus. The volume of each shape schematically represents the structural thickness required to achieve the corresponding absorption spectrum. The brown and blue shapes correspond to the conventional isothermal modulus, $B_{\mathrm{iso}}$, and the sorption-modified modulus, $B_{\mathrm{sorp}}$, respectively. For the same absorption spectrum, reducing the static bulk modulus decreases the volume of the shape, indicating a lower required thickness.
(b) Normal-incidence sound absorption coefficient of a $40\,\mathrm{mm}$-thick granular activated carbon (GAC) absorber. Open circles represent measurements, and the solid blue line represents the theoretical prediction including sorption.
(c) Causality-based minimum thickness calculated from the same absorption spectrum as a function of the upper frequency limit of integration. The brown and blue curves correspond to the conventional isothermal and sorption-modified static moduli, respectively. Horizontal dashed lines indicate the corresponding full-spectrum minimum thicknesses of $69.61\,\mathrm{mm}$ and $39.64\,\mathrm{mm}$.}
\label{fig:3}
\end{figure}

The experimentally demonstrated reduction in static modulus adds a constitutive design axis to the conventional frequency--response design space. Alongside shaping how absorption is distributed over frequency, absorber design can therefore incorporate the static constitutive response as an additional material degree of freedom (Fig.~\ref{fig:3}(a)).

To examine how this additional design freedom manifests at the device level, we considered a 40-mm-thick GAC absorber under normal incidence (see Supplemental Material, Section V.A for detailed input parameters of the GAC). The measured absorption spectrum agrees well with the theoretical prediction over 100--1600~Hz (Fig.~\ref{fig:3}(b)), demonstrating that the constitutive model captures the acoustic response of the finite absorber. Applying the causal relation to the same absorption spectrum gives a minimum thickness of 69.61~mm under the conventional isothermal reference, whereas the sorption-modified static modulus reduces this value to 39.64~mm (Fig.~\ref{fig:3}(c)). The physical thickness of 40~mm is therefore consistent with the sorption-modified causal prediction while remaining substantially below the conventional isothermal benchmark. This quantitative agreement demonstrates how constitutive control of the static modulus translates directly into a reduced thickness requirement for broadband absorption.

\FloatBarrier
\section{Spectral engineering of sorption-enabled broadband absorbers}

\begin{figure}[!htbp]\centering
\includegraphics[width=89mm,keepaspectratio]{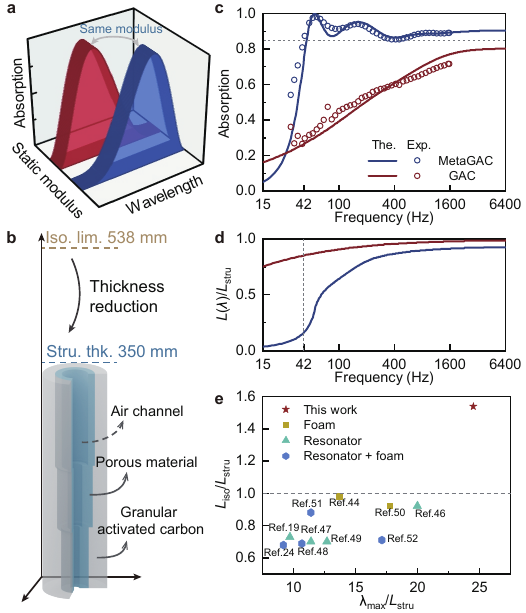}
\caption{\textbf{Spectral engineering of sorption-enabled broadband sound absorption.}
(a) Schematic illustration of absorption-spectrum tuning at the same structural thickness and static bulk modulus.
(b) Cutaway schematic of MetaGAC, which combines granular activated carbon (GAC), melamine foam, and air channels in a cascaded hierarchical structure. Horizontal dashed lines mark the physical thickness of $350\,\mathrm{mm}$ and the conventional isothermal causal thickness benchmark of $538\,\mathrm{mm}$ for the corresponding absorption spectrum.
(c) Normal-incidence absorption spectra of MetaGAC and homogeneous GAC with the same overall thickness. Blue and red denote MetaGAC and GAC, respectively; solid lines show theoretical predictions, and open circles show measurements. The horizontal dashed line marks an absorption coefficient of $0.85$.
(d) Cumulative causal thickness normalized by the physical thickness, $L(f)/L_{\mathrm{stru}}$, plotted against the upper frequency limit of integration. Blue and red curves represent MetaGAC and homogeneous GAC, respectively. The vertical dashed line marks $42\,\mathrm{Hz}$, below which MetaGAC exhibits a substantially smaller cumulative contribution to the causal thickness.
(e) Performance comparison between MetaGAC developed in this work and representative previously reported broadband absorbers, including foams, resonators, and resonator--foam composites. The horizontal and vertical axes show the wavelength-to-thickness ratio, $\lambda_{\max}/L_{\mathrm{stru}}$, and the isothermal causal-thickness ratio, $L_{\mathrm{iso}}/L_{\mathrm{stru}}$, respectively. Here, $\lambda_{\max}$ is the wavelength at the lower edge of the continuous absorption band defined by an absorption coefficient of at least 0.8, $L_{\mathrm{stru}}$ is the physical thickness, and $L_{\mathrm{iso}}$ is the conventional isothermal causal minimum thickness for the corresponding absorption spectrum. The red star denotes this work. The horizontal dashed line marks $L_{\mathrm{iso}}/L_{\mathrm{stru}}=1$, indicating the causal thickness limit within the conventional nonsorbing isothermal framework.}
\label{fig:4}
\end{figure}

The reduced static modulus established for GAC lowers the causal thickness requirement. Whether this constitutive advantage is used efficiently, however, depends on how the absorption is distributed across frequency. In a homogeneous GAC layer, the frequency-dependent response is constrained by its transport characteristics, including relatively high flow resistance\cite{Venegas-2016-JASA} (see detailed analysis in Supplemental Material, Sec. V.B), and the resulting absorption is not necessarily concentrated within the desired frequency range. Because absorption outside the target band also contributes to the causal integral, off-band response increases the associated thickness requirement without contributing to the intended in-band performance\cite{Yang-2017-MH,Meng-2022-PRSA}. Efficient use of the reduced causal thickness therefore depends on redistributing absorption toward the frequency range of interest (Fig.~\ref{fig:4}(a)).

Melamine foam provides efficient porous dissipation, particularly at higher frequencies\cite{Allard-2009-Book,Huang-2023-PRA}, while previous studies of double-porosity absorbers have shown that combining distinct pore domains through structural design can broaden and reshape acoustic absorption. Building on this principle, we integrate sorptive GAC with melamine foam and air channels through a deliberately designed cascaded hierarchical architecture. Following the metamaterial strategy of combining constituent properties with structural degrees of freedom, we term the resulting composite MetaGAC (Fig.~\ref{fig:4}(b)). Compared with homogeneous GAC of the same overall thickness, MetaGAC exhibits a substantially reshaped broadband response. The measured absorption spectrum agrees well with the theoretical prediction, with the measured absorption coefficient remaining above 0.85 from $42\,\mathrm{Hz}$ to $1600\,\mathrm{Hz}$ (Fig.~\ref{fig:4}(c); the theoretical analysis and experiment measurement are detailed in Supplemental Material, Sec. VI and VII). The theoretical prediction further yields an average absorption coefficient of 0.91 over $40$--$6400\,\mathrm{Hz}$. The MetaGAC structure has a physical thickness of 350 mm, whereas the conventional isothermal framework gives a causal thickness benchmark of 538 mm for the corresponding absorption spectrum, demonstrating that the broadband response is achieved at a substantially reduced structural thickness.

The origin of this improvement becomes clearer when the absorption spectra are examined through their cumulative contribution to the causal thickness. Because the causal relation is integrated over wavelength, even modest absorption at very low frequencies can contribute disproportionately to the required thickness owing to the long wavelengths involved. For homogeneous GAC, more than 85\% of the normalized causal-thickness contribution accumulates below 42 Hz, although this region lies outside the target absorption band. MetaGAC strongly suppresses this off-band contribution, reducing the corresponding fraction to approximately 15\% (Fig.~\ref{fig:4}(d)). The causal-thickness contribution is thereby redistributed toward the target frequency range, allowing the structural thickness to be utilized more efficiently for the desired broadband response.

The distinction becomes particularly evident when MetaGAC is compared with representative broadband absorbers in a multidimensional performance space (Fig.~4e)\cite{Wang-2025-NSR,Qu-2022-SA,Zhou-2022-NSR,Yang-2017-MH,Ding-2022-IJMS,Mei-2024-IJMS,Wang-2025-IJMS,Ge-2025-PRL,Qian-2024-IJMS,Wang-2024-MD}. Of the two metrics considered, the isothermal causal-thickness ratio, $R_{\mathrm{ICT}}={L_{\mathrm{iso}}}/{L_{\mathrm{stru}}}$,
directly reflects the constitutive advantage developed in this work. Within the conventional isothermal framework, $L_{\mathrm{iso}}$ represents the minimum thickness based on causality required for the corresponding absorption spectrum, such that $R_{\mathrm{ICT}}\leq 1$. Sorption reduces the effective static bulk modulus and consequently lowers the causal minimum thickness associated with the same absorption spectrum, allowing the physical structure to be thinner than the conventional isothermal benchmark and hence enabling $R_{\mathrm{ICT}}>1$. MetaGAC reaches $R_{\mathrm{ICT}}\approx1.538$, placing it beyond the conventional isothermal thickness reference.

Alongside this thickness advantage, MetaGAC achieves a large wavelength-to-thickness ratio, $R_{\mathrm{WT}}=\lambda_{\max}/L_{\mathrm{stru}}=24.5$, where $\lambda_{\max}$ is the wavelength at the lowest frequency above which the absorption coefficient remains at or above 0.8 throughout the evaluated range. MetaGAC thus combines broadband absorption with low-frequency compactness and a thickness below the conventional isothermal causal benchmark.

\FloatBarrier
\section{Conclusion}

In summary, we show that microscopic gas-surface processes can provide an additional degree of freedom for material design, with sorption under micro- and nanoscale confinement modifying the equilibrium acoustic constitutive response. Theoretical analysis shows that adsorption introduces additional equilibrium compressibility, allowing the static bulk modulus to fall below the conventional isothermal reference and thereby reducing the causality-based minimum thickness required for a prescribed absorption spectrum. We experimentally validate this mechanism using granular activated carbon (GAC), for which low-frequency bulk-modulus measurements agree well with the sorption-inclusive model and yield, through extrapolation to the zero-frequency limit, a normalized effective static bulk modulus of 0.567. Building on this constitutive response, we further combine GAC with spectral and structural engineering to realize MetaGAC, which maintains an absorption coefficient above 0.85 from 42~Hz with a physical thickness of 350~mm, substantially below the conventional isothermal causal thickness benchmark of 538~mm for the corresponding absorption spectrum. These results establish a quantitative connection between microscopic sorption at internal interfaces, the macroscopic equilibrium acoustic constitutive response, and the causal thickness requirement of the absorber. More broadly, this work points to a materials-design strategy in which microscopic and interfacial physical processes can be exploited as degrees of freedom for engineering the constitutive properties that ultimately constrain macroscopic performance.

\FloatBarrier
\bibliography{references_merged}
\end{document}